# Hydrogen bond relaxation dynamics and the associated vibronic and volumetric anomalies of H$_2$O upon frozen

Xi Zhang,[1] Xiaojian Fu,[2] Yu Wen,[3] Jer-lai Kuo,[4] Zexiang Shen,[5] Ji Zhou,[2*] Chang Q Sun[1,3*]

[1] School of Electrical and Electronic Engineering, Nanyang Technological University, Singapore 639798
[2] State Key Laboratory of New Ceramics and Fine Processing, Department of Materials Science and Engineering, Tsinghua University, Beijing 100084, China
[3] Faculty of Materials and Optoelectronic Physics, Xiangtan University, Hunan 411105, China
[4] Institute of Atomic and Molecular Sciences, Academia Sinica, Taipei 10617, Taiwan
[5] School of Mathematical and Physical Science, Nanyang Technological University, Singapore 639798
***E-mail:** Zhouji@tsinghua.edu.cn; Ecqsun@ntu.edu.sg

A combination of the extended Ice Rule of Pauling, Raman spectroscopy, and molecular dynamics calculations has enabled us to clarify the bonding origin of the anomalous volume expansion, Raman phonon relaxation, and the stiffness and fragility of H$_2$O upon frozen. We found that the initially shorter-and-stronger intramolecular "H$^{+/p}$-O$^{2-}$" bond follows the regular rule of cooling-contraction while the initially longer-and-weaker intermolecular "O$^{2-}$ : H$^{+/p}$" nonbond turns to be even longer yet stiffer in the "O$^{2-}$ : H$^{+/p}$-O$^{2-}$" hydrogen-bond of H$_2$O upon frozen, as a consequence of the polarization and Coulomb repulsion between the unevenly-bounded bonding and nonbonding electron pairs. The elongation of the nonbond and the polarization of the nonbonding lone pair are responsible, respectively, for the volume expansion and the stiffness and the fragility of ice. Findings should form important impact to the understanding of the physical anomalies of H$_2$O under other stimuli such as pressure and confinement.

As the central element of the universe and living species such as DNA folding,[1, 2] unfolding,[3] and messaging,[4] protein[5] and gene delivery,[6, 7] dye sensitization[8], drug molecule target binding,[9] $H_2O$, and the hydrogen bond as well, is too anomalous, too strange, and too challenge.[10] Although the structure formation and phase transition[11, 12, 13] and the reaction dynamics[14, 15, 16] of $H_2O$ have been intensively investigated, the origin of its physical anomalies has been continuing puzzling the community for generations.[17, 18, 19, 20, 21, 22, 23] For instances, it is usual for liquids to contract on freezing and expand on melting because the molecules are in fixed positions within the solid but require more space to move around within the liquid. When water freezes at 0°C its volume increases by up to 9% under atmospheric pressure compared with liquid argon that shrinks by 12% on freezing.[24] Unlike conventionally known materials, ice is stiff yet fragile. These anomalies is beyond the expectation of Pauling's Ice Rule[25] that predicted the crystal geometry and entropy of various phases.

Although intensive investigation has been conducted on the vibronic behavior of $H_2O$ and the hydrogen bond using Raman and Infrared spectroscopies under various conditions,[26, 27, 28, 29, 30, 31, 32] correlation between the hydrogen bond relaxation dynamics and the observed vibronic and volumetric anomalies of $H_2O$ upon frozen needs to be established. For instances, Durickovic et al[27] measured recently the high-frequency O-H stretching mode within the temperature range of 10 and -15 °C and found phase transition happens at 0 ± 3 °C associated with redshift of the high-frequency mode upon ice formation. Pruzan at al[29] found that redshift happens to the high-frequency mode while the blueshift to low-frequency vibration mode when the applied pressure and the measuring temperature is increased. These observations are beyond the scope of expectations that cooling and compressing induce stiffening of all the possible phonon modes like carbon allotropes.[33]

The aim of this communication is to show that a combination of the extended Ice Rule of Pauling,[25] programmable Raman spectroscopy, and molecular dynamics (MD) calculations has enabled us to clarify and correlate these concerns leading to comprehension of the origin of the vibronic and volumetric anomalies of $H_2O$ upon frozen from the perspective of the real and virtual bond relaxation dynamics upon frozen.

Figure 1 (a) shows the extension of Pauling's Ice Rule[25] (the central tetrahedron) to contain two $H_2O$ molecules and four identical quasi-linear "$O^{2-}$ : $H^{+/p}$-$O^{2-}$" hydrogen bonds.[34] The $H^{+/p}$ plays a dual role of H$^+$ and H$^P$ as it donates its electron to one O$^{2-}$ and meanwhile it is polarized by the nonbonding lone pair of the other neighboring O$^{2-}$ upon the sp-orbit of oxygen being hybridized in reaction.[35] In the hexagonal or cubic ice the O$^{2-}\cdots$O$^{2-}$ distance is 0.276 nm. The intramolecular H$^{+/p}$-O$^{2-}$ bond is much shorter and stronger (~0.100 nm and ~10$^0$ eV) than that of the intermolecular O$^{2-}$ : H$^{+/p}$ nonbond (~0.176 nm and ~10$^{-2}$ eV). The angle between the H$^{+/p}$–O$^{2-}$–H$^{+/p}$ is smaller than 104.5° while the angle between the H$^{+/p}$ : O$^{2-}$ : H$^{+/p}$ is greater than 108.5° for a free H$_2$O molecule. It has recently been confirmed[36] that the liquid water is consistent with a unimodel density of the tetrahedral structure at the ambient conditions. The advantage of the extended Ice Rule is that it allows us to focus on one of the four "O$^{2-}$ : H$^{+/p}$-O$^{2-}$" bonds to examine the responses of the intramolecular nonbonding lone pair ":" and the intermolecular bonding pair "–" of electrons to the applied stimulus and their correlation separately.

Figure 1 (b) shows the expected relaxation dynamics of the two partitions. If the real bond follows the regular rule of cooling expansion, the virtual nonbond will expand because of the Coulomb repulsion between these two unevenly-bounded electron pairs. The net gain of the O$^{2-}$---O$^{2-}$ length is responsible for the volume expansion of H$_2$O upon frozen. The polarization and the extremely weak interaction make the nonbond stiffer and fragile; The stiffening of the virtual bond will gives rise to the blue shift of the soft Raman mode; the shortening of the real bond will lead to the blue shift of the stiff Raman mode according to convention that the shorter bond becomes stronger.

In order to verify the hypotheses based on the extended Ice Rule, we conducted Raman measurements and MD computations of H$_2$O in the temperature range cross the frozen point. The Raman spectroscopy is the unique tool that it could resolve the vibration frequencies of the two partitions of the intramolecular nonbond in the frequency range of $\omega_L < 300$ cm$^{-1}$ and that of the intermolecular bond in the frequency range of $\omega_H > 3000$ cm$^{-1}$. From the first order approximation and conventional approach for other existing materials,[37] the bonding and nonbonding part of the hydrogen bond can be taken as each a harmonic system with an interaction potential, u(r$_x$), with subscript x representing the ":" and "-" partitions. Equaling the vibration energy of the harmonic system to the third term of the Taylor series of its interaction potential at equilibrium, we can obtain the relation:[38]

$$\frac{1}{2}\mu(\Delta\omega_x)^2(r-d_x)^2 \cong \frac{1}{2}\left.\frac{\partial u(r)}{\partial r^2}\right|_{r=d_x} x^2 \propto \frac{1}{2}\frac{E_x}{\mu d_x^2}(r-d_x)^2$$

$$\Delta\omega_x = \omega_x - \omega_{x0} \propto \frac{E_x^{1/2}}{\mu d_x}$$

(1)

Generally, the Raman shift depends on the length and energy of the bond and the reduced mass of the atom or molecule of the vibronic system. From the dimensional point of view, the second order derivative of the potential at equilibrium is proportional to the bind energy $E_x$ divided by the bond length in the form of $d_x^2$. The $E_x^{1/2}/d_x \cong \sqrt{Y_x d_x}; (Y_x \approx E_x/d_x^3)$ is right the square root of the stiffness being the product of the Young's modulus and the bond length.[38] $\omega_{x0}$ is the referential point from which Raman shift proceeds. This relation indicates that if the bond or the nonbond contracts or the binding energy increases, or the bond becomes stiffer, a blue shift will happen. Therefore, Raman frequency shift is able to tell us the change of the length and energy of the respective partition of the hydrogen bond. This approach has been successfully used in analyzing the Raman shift of Si, C, and oxides, for which the bond energy is inversely proportional to a certain power of the bond length.[37, 38] Generally, a bond becomes weaker if it becomes longer and the bond turns to be stronger when it becomes shorter. This tradition keeps no valid for the bisectors of the hydrogen bond of $H_2O$ because of the involvement of the repulsion between the unevenly-bounded bonding and nonbonding electron pairs.

Raman measurements of 0.5 gram deionized water poured on silica stage were conducted using the Lab RAM HR800 Raman spectrometer (HORIBA Jobin Yvon Ltd) with the 632.8 nm He-Ne laser as light source. The semiconductor refrigeration cooling detection systems were used to collect the data under the programmed controlling temperature and at the ambient pressure. The frequency range was set at 50-1000 and 2600-4000 cm$^{-1}$. The temperature was lowered from 298 K to 98 K in 3K/min rate and 10 K step size (some 5 K step size). The spectrum is an accumulation of 4 scans and each scan took 30 sec. At each step, the measurement was conducted after the temperature being held for 5 minutes. The measurements were repeated in the processes of temperature rise and drop and conducted at both Tsinghua University (full temperature and frequency range) and at Nanyang Technological University (high frequency at 298 and 248 K) to ensure the accuracy.

The temperature-resolved Raman spectra in the specified frequency range of $\omega_L < 350$ cm$^{-1}$ and $\omega_H > 3000$ cm$^{-1}$, in Figure 2, shows the following:

1. Solid-liquid phase transition happens at T ≥ 268 K. Other phase transitions may happen in the

solid phase (dramatic change in the Raman spectral shape), but that is not the immediate concern of this presentation.

2. At T > 271 K, no substantial frequency-peak change happens to the $\omega_H$ because the heating energy (meVs) is insufficient to excite relaxation of the stronger bond with energy of eVs; however, the $\omega_L$ is sensitive to the temperature with an additional hump presented to the $\omega_L$ at 175 cm$^{-1}$ in addition to the original peak at 75 cm$^{-1}$ when the temperature is dropped from 298 to 288 K. This finding suggests that one should focus on the $\omega_L$ as the finger print of the hydrogen bond relaxation in the liquid phase instead of the $\omega_H$.

3. At the water-ice transition point, T ~ 268 K, the $\omega_H$ shifts suddenly from 3200 to 3150 cm$^{-1}$ and the $\omega_L$ shifts completely from 75 to 220 cm$^{-1}$. The ice formation is supposed to happen at 273 K, the additional pressure upon ice formation will actually lowers the melting point,[39] a ± 5 K error bar does not affect the conclusion made.

4. With further cooling, the $\omega_H$ shifts gradually from 3150 cm$^{-1}$ to 3100 cm$^{-1}$ and the $\omega_L$ shifts from 220 to 230 cm$^{-1}$, disregarding other supplementary peaks at ~300 and ~3400 cm$^{-1}$.

All observations are within the expectation but the redshift of the stiff Raman mode that is abnormal. Incorporating the analytical expression to the measurements suggest that at the point of solid-liquid phase transition, the weak intermolecular $O^{2-}$ : $H^{+/p}$ nonbond becomes stiffer; the strong intramolecular $H^{+/p}$-$O^{2-}$ bond becomes softer. According to the conventional perception, the $H^{+/p}$-$O^{2-}$ becomes longer and weaker while the $O^{2-}$ : $H^{+/p}$ becomes shorter and stronger upon frozen. However, this transition is absolutely not the case of ice formation associated with 9% volume expansion.

The volume relaxation of 96 water molecules with disordered protons from 250 to 277K was then calculated using the Forcite package[40] of MD with *ab initio* optimized force field Compass27. The structures were relaxed in the Isothermal-Isobaric ensemble for 30ps, and then in microcanonical ensemble for 10ps, at a time interval of 0.5 fs each step. The average $O^{2-}$-$H^{+/p}$ and $O^{2-}$ : $H^{+/p}$ lengths were taken of the structures of last 10ps (20,000 steps). Table 1 lists and Figure 3 shows the length correlation between the $O^{2-}$ : $H^{+/p}$ and the $H^{+/p}$-$O^{2-}$. The $H^{+/p}$-$O^{2-}$ bond contracts from 0.0979 to 0.0971 nm and meanwhile the $H^{+/p}$ : $O^{2-}$ nonbond expands from 0.1878 to 0.1942 nm with a net linear expansion of ~2.0%, or 6% volume expansion, approaching fairly the known volume expansion of ice but the current calculation is based on water instead of ice.

Therefore, we have to separate the binding energy and the bond length in eq (1) for the two partitions of the H-bond with consideration of the effect of polarization and the screening of the bonding potential by the neighboring dipoles.[34] The unexpected inconsistency in the length and Raman stiffness suggested that the real bond is indeed shortened but softened because of the screening of the potential by the lone-pair induced neighboring dipoles. The stiffening of the soft Raman mode is correlated to the polarization, which is responsible for the stiffness and the fragility ice. Hence, the hydrogen bond relaxation dynamics, the Raman phonon relaxation, the volumetric and vibronic anomalies of $H_2O$ upon frozen have been correlated and understood.

In summary, combining the Raman spectroscopy and MD calculation, we are certain that the initially shorter and stronger $H^{+/p}$-$O^{2-}$ bond follows the regular rule of cooling contraction but it becomes weaker because of the screening effect; the initially longer and weaker $O^{2-}$ : $H^{+/p}$ becomes even longer but stiffer upon frozen. The transition happens because of the repulsion between the unevenly-bounded electron pairs and the polarization of the nonbonding lone pair by the electrons of the cooling-contracted real bond that is in turn screened by the former. Practice and findings may pave the path towards controlling the hydrogen-bond formation and relaxation dynamics and find applications in the hydrogen-bond involved species such as drug design, medication, signaling and messaging.

Financial support from NSF (Nos.:1033003, 90922025 and 11172254) China is gratefully acknowledged.

Table and Figure captions

Table 1 MD derived lengths of the two parts and the resultant of the hydrogen bond of water.

| T(K) | O-H (Å) | O:H (Å) | O-O(Å) |
|---|---|---|---|
| 277 | 0.9791 | 1.8797 | 2.8588 |
| 273 | 0.9764 | 1.8961 | 2.8726 |
| 270 | 0.9740 | 1.9084 | 2.8824 |
| 260 | 0.9715 | 1.9252 | 2.8967 |
| 250 | 0.9709 | 1.9399 | 2.9108 |

Figure 1 (a) the extension ($2H_2O$ tetrahedron) of Pauling's "two-in two-out" ice rule (the central $H_2O$ tetrahedron) allows us to focus on the cooperative interaction between the unevenly-bounded nonbonding lone pair ":" and bonding pair "–" of electrons in the quasi-linear "$O^{2-}$ : $H^{+/p}$-$O^{2-}$" hydrogen bond. (b) The real bond follows the regular rule of cooling contraction but the virtual bond is elongated by the repulsion between the unevenly bonded bonding and nonbonding electron pairs. The polarization makes the virtual bond stiffer and the weak nonbond makes ice fragile.

Figure 2 Temperature-resolved Raman shift of (a) the intramolecular nonbonding lone pair $O^{2-}$ : $H^{+/p}$ ($\omega_L$ < 350 cm$^{-1}$) and (b) the intermolecular bonding pair $O^{2-}$-$H^{+/p}$ ($\omega_H$ > 3000 cm$^{-1}$) shows the melting point at T ~ 268 K. Despite the phase transitions, the blueshift of the soft mode (a) indicates the stiffening of the nonbond and the redshift of the stiff mode (b) indicates the softening of the bond.

Figure 3 MD derived correlation of the $O^{2-}$-$H^{+/p}$ and the $O^{2-}$ : $H^{+/p}$ lengths.

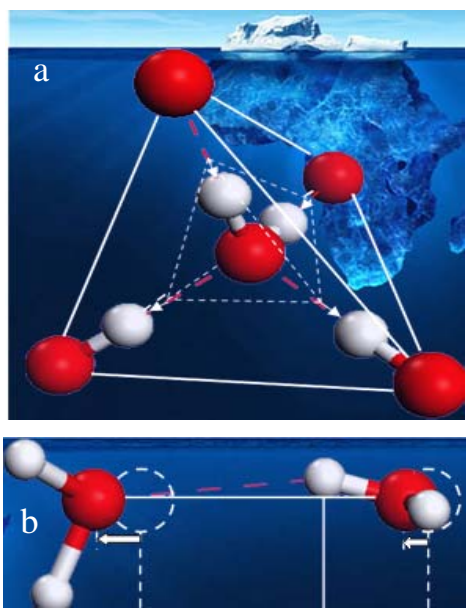

Figure 1

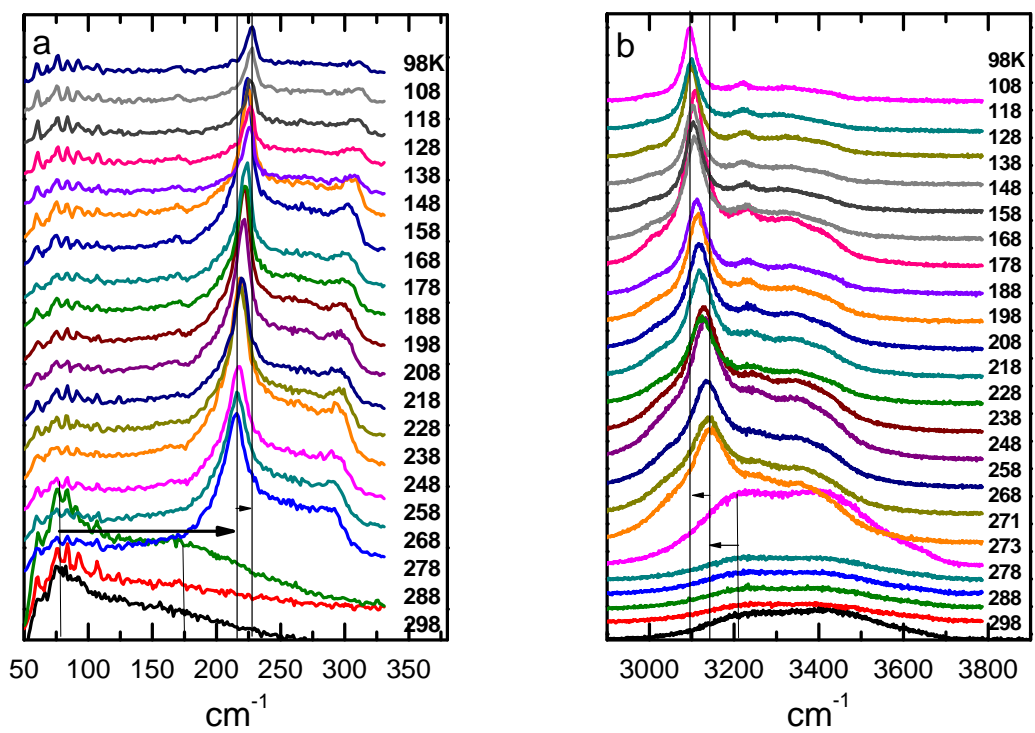

Figure 2

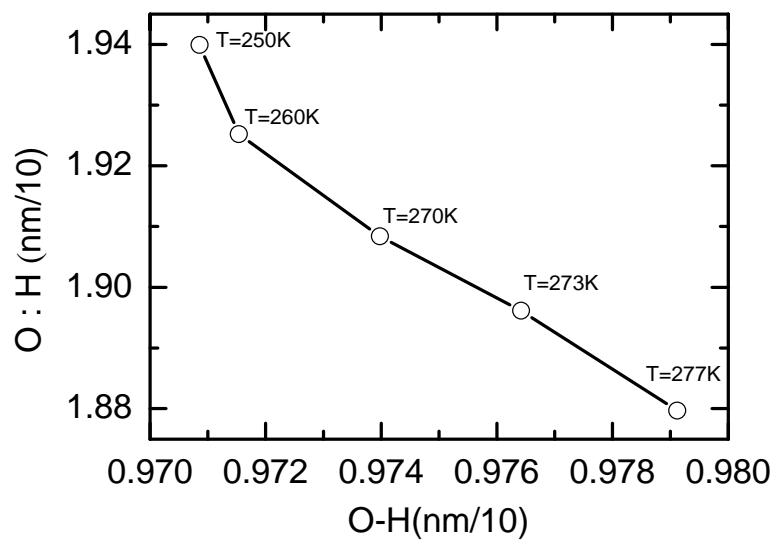

Figure 3